\documentclass[rotate]{an_art}
\usepackage{lscape}
\usepackage{longtable}
\def\p1leiste #1 #2 #3 {{\noindent\Large Volume #1 \hfill {#2}
\hfill {Number #3}
\vspace{2mm}\par\hrule\par\vspace{8mm}}}
\def\an #1 #2 #3 #4 #5 {{\noindent\small Astron.~Nachr.~#1 (#2) #3, #4--#5
\bigskip\par}}

\begin{document}

\begin{titlepage}
\setcounter{page}{001}
\headnote{Astron.~Nachr.~321 (2000) 1, 1--52}
\def\makeheadline{\vbox to 0pt{\vskip -30pt
\hbox to 174mm{\huge Astronomische  Nachrichten \hfill}}}
\makeheadline
\p1leiste 321 2000  1
\an 321 2000 1 1 52
\title {The ROSAT Bright Survey:\\ II. Catalogue
of all high-galactic latitude RASS sources \\with PSPC countrate  
CR $> 0.2$\,s$^{-1}$\thanks{Based in part on observations at the 
European Southern Observatory
La Silla (Chile) with the $\rm 2.2m$ telescope of the Max-Planck-Society}}

\author{{\sc A.D.~Schwope, G. Hasinger, I. Lehmann, R. Schwarz, H. Brunner,}
Potsdam, Germany\\
\medskip
{\small Astrophysikalisches Institut Potsdam} \\
\bigskip
 {\sc S. Neizvestny, A. Ugryumov, Yu.~Balega}, Nizhnij Arkhyz, Russia \\
 \medskip
 {\small Special Astrophysical Observatory}\\
\bigskip
 {\sc J.~Tr\"{u}mper, W.~Voges}, Garching, Germany \\
 \medskip
 {\small Max-Planck-Institut f\"ur extraterrestrische Physik}
}

\date{Received 2000 January 28; accepted 2000 February 7} 
\maketitle

\summary
We present a summary of an identification program of 
the more than 2000 X-ray sources
detected during the ROSAT All-Sky Survey (Voges et al.~1999) 
at high galactic latitude, $|b| > 30\degr$,
with countrate above 0.2\,s$^{-1}$. This program, termed the ROSAT 
Bright Survey RBS, is to more than 99.5\% complete. 
A sub-sample of 931 sources with countrate above
0.2 s$^{-1}$ in the hard spectral band between 0.5 and 2.0 keV is 
to 100\% identified. 
The total survey area comprises 20391 deg$^2$ 
at a flux limit of $2.4 \times 10^{-12}$\,erg cm$^{-2}$ 
s$^{-1}$ in the 0.5 -- 2.0 keV band. 
About 1500 sources of the complete sample could be 
identified by correlating the RBS with SIMBAD and the NED.
The remaining $\sim$500 sources were identified 
by low-resolution optical spectroscopy and CCD imaging 
utilizing telescopes at La Silla, 
Calar Alto, Zelenchukskaya and Mauna Kea. 
Apart from completely untouched sources, 
catalogued clusters and galaxies without published redshift as well as
catalogued galaxies with 
unusual high X-ray luminosity were 
included in the spectroscopic identification program. 
Details of the observations with 
an on-line presentation  of the finding charts and the optical spectra 
will be published separately. Here we summarize our 
identifications in a table which contains optical and X-ray 
information for each source. 
As a result
we present the most massive complete sample of X-ray selected AGNs with 
a total of 669 members and a well populated X-ray selected sample of 
302 clusters of galaxies
with redshifts up to 0.70. Three fields studied by us remain 
without optical counterpart (RBS0378, RBS1223, RBS1556). While the first is a 
possible X-ray transient, the two latter are 
isolated neutron star candidates 
(Motch et al.~1999, Schwope et al.~1999).END

\end{titlepage}

\section{Introduction}
A decade after launch of the ROSAT satellite, narrow and wide angle surveys
using All-Sky Survey (RASS) data or pointed observations become complete. 
In order to characterize the source content of the X-ray sky as seen by
ROSAT, such surveys were performed 
at high galactic latitudes by different teams. The most important 
surveys based on pointed observations are 
the deep surveys in the Lockman Hole (Hasinger et al.~1998, 
Schmidt et al.~1998), the 
UK deep survey (McHardy et al.~1998), and the 
medium deep RIXOS survey (Mason et al.~2000). Appenzeller et al.~(1998)
presented a catalogue of sources in selected areas of the sky 
down to the RASS limit.
Thomas et al.~(1997) performed a high-galactic latitude survey ($|b| > 
20\degr$) of bright, point-like, soft sources (HR1 + $\Delta$HR1 $ < $ 0.0), 
the corresponding survey of bright, point-like, hard sources (HR1 $ > $ 0.5)
was presented by Fischer et al.~(1998, referred to a Paper I). 
For definition of the hardness 
ratio HR1 see Sect.~\ref{s:tab} Based on their Schmidt plate survey run at 
Calar Alto, Bade et al.~(1995, 1998) published a catalogue of northern 
AGNs detected in the RASS. The missing piece, an unbiased survey 
of all bright, CR $ > 0.2$\,s$^{-1}$, 
high-galactic latitude sources detected in the RASS without
further selection, e.g.~for X-ray extent or X-ray color, is presented here.
It is termed the ROSAT Bright Survey RBS.

The survey area of the RBS
is the sky above galactic latitude $30\degr$ and below $-30\degr$ excluding
the Virgo cluster and the Magellanic clouds. 
The 1RXS-catalogue by Voges et al.~(1999) contains 2072 high-galactic 
latitude sources 
brighter than our count rate limit.
We refer to these as RBS0001 \dots  RBS2072
with increasing rightascension. After exclusion of the Virgo and MC regions,
approximated as circles with radius $5\degr$ centred on $(\alpha,\delta) = 
(16\fd44,-73\fd27), (83\fd80,-68\fd00)$, and $(188\fd30,1\fd70)$ the remaining
survey area is 20391 deg$^2$ and contains 2012 sources. 
These were identified almost completely by catalogue work and a dedicated
identification program. One of the main results of this survey is a 
catalogue of all 2012 sources summarizing their main properties at X-ray and
optical wavelengths which is presented in this paper. A description 
of the source content highlighting the extremes and the statistics 
of the survey is in 
preparation and will be published in the near future. At that time the 
catalogue will go online, i.e.~finding charts and identification 
spectra will become electronically accessable. 

One of the major results of the RBS 
was the determination of the 
soft X-ray AGN luminosity function and, 
combined with deeper surveys, the assessment of its 
cosmological evolution and its contribution to the soft X-ray 
background (Miyaji et al.~2000). Some of the new interesting galactic 
sources found in the RBS, magnetic cataclysmic variables and isolated
neutron stars, were presented in Schwope et al.~(1997, 1999a, 1999b) 
and Motch et al.~(1999).

\section{Description of the RBS-catalogue}
\label{s:tab}
The table appended below 
gives summarizing information about all 2072 RBS-sources. It 
is organized in 17 columns, which are described subsequently:\\
{\it (1)} The RBS sequence number\\
{\it (2)} The entry in the 1RXS-catalogue (X-ray coordinates for equinox 
J2000)\\
{\it (3,4)} The optical coordinates of the X-ray source for equinox J2000\\
{\it (5)} The distance between the optical and the X-ray coordinate 
in arcsec \\
{\it (6)} The X-ray positional uncertainty ($1\sigma$) 
in arcsec as given in the 1RXS-catalogue. 
The 90\% confidence error radius is 
$r_{90} = 1.65\sigma$\\
{\it (7)} The RASS countrate in the total ROSAT window 0.1--2.4 keV\\
{\it (8)} The RASS countrate in the hard ROSAT window 0.5--2.0 keV\\
{\it (9)} The hardness ratio HR1=(H+S)/(H--S), with H and S being the 
counts in the hard (0.5--2.0 keV) and the soft (0.1--0.4 keV) spectral bands,
respectively\\
{\it (10)} An alternate name of the source (for sources with catalogue 
entries others than the RBS), truncated to 10 characters\\
{\it (11)} The class and the type of the X-ray source 
with the following abbreviations (a colon indicates an uncertain
identification):\\
\begin{longtable}{lp{7cm}p{8cm}}
Class &  Explanation & Type\\[2mm] \hline
\ \\
\endfirsthead 
Class &  Explanation & Type\\[2mm] \hline
\ \\
\endhead
\ \\ 
\hline\ \\
\endfoot
\ \\
\hline\ \\
\endlastfoot
AGN & active galactic nucleus, X-rays originate from the central engine &
        QSO, BLL (BL Lac object),
 	Sy1\dots2 (Seyfert-Galaxy of subtype 1\dots2), 
	NLS1 (Narrow Line Seyfert 1 galaxy), LINER, 
	XTG (X-ray transient galaxy), NELG (narrow emission line galaxy),
	SRB (starburst galaxy)\\
GALAX & normal galaxy without obvious nuclear activity. X-rays originate 
	predominantly from the stellar constituents of the galaxy &
	 Hubble type \\
CLUST & cluster of galaxies  & no further division\\
GGRP  & small (compact) group of galaxies & no further division\\
STAR  & coronal or photospheric emitter of X-rays & 
	PN (planetary nebula), spectral type (for coronal emitters and 
	hot stars), WD (white dwarf),
	INS (isolated neutron star)\\
CV & cataclysmic variable or related object (white dwarf accretor)&
 	AM (AM Herculis star or polar), IP (intermediate polar), 
	DN (dwarf nova), NL (novalike variable), SSS (supersoft source), 
	Symb (symbiotic binary)\\
XRB & X-ray binary (neutron star or black hole accretor) &  
	HMXB (high-mass X-ray binary), LMXB (low-mass X-ray 
	binary), XRT (X-ray transient)\\
MC & Magellanic cloud source, not considered in the identification 
program\\
VIRGO & X-ray source in the Virgo region,  
not considered in the identification program\\
\end{longtable}
Plausible identifications based on NED or SIMBAD correlation 
entered the table without checking the original literature. If the  
class or type of a specific source seemed questionable for its specific
properties (e.g. unusual high X-ray flux
or X-ray extent), it entered our spectroscopic identification program. \\
{\it (13)} The redshift of the X-ray source (if extragalactic), a colon 
indicates an uncertain value.\\
{\it (14)} An optical magnitude of the X-ray source, a colon indicates 
an uncertain value. The character behind the value identifies the source
of information: C -- CDS/SIMBAD (V-band), D -- CDS/SIMBAD (B-band), N --
NED (a magnitude in the optical range), R -- APM/ROE scan of POSS/ESO/SERC 
red (E) plate, B -- APM/ROE scan of POSS/ESO/SERC blue (O) plate, S -- 
below slit magnitude determined by folding a low-resolution spectrum 
with the sensitivity curve of a V-filter\\
{\it (15)} The X-ray flux in the 0.5--2.0 keV range in units 
of $10^{-14}$\,erg cm$^{-2}$ s$^{-1}$. 
Assuming a power law spectrum with photon index 2 a count-to-energy 
conversion factor ECF for the ROSAT PSPC was 
computed  for given column density towards the X-ray source (galactic 
$N_{\rm H}$ assumed). The flux was then computed using the 
countrate in the hard band (column 8) via $f_x =$ CRH/ECF. This approach
yields unreliable and inconsistent 
results for sources with HR1 smaller than --0.4, even an unabsorbed power law
cannot be softer than this limit.\\
{\it (16)} The log of the X-ray luminosity in ergs/s computed as 
$L_x = 4 \pi f_x (cz/H_0)^2$, with $H_0 = 50$ km s$^{-1}$ Mpc$^{-1}$.\\
{\it (17)} A hint to a note or comment about the source appended below 
the table.\\

\acknowledgements
The ROSAT project is supported by the Bundesministerium f\"{u}r 
Bildung, Wissenschaft, Forschung und Technologie (BMBF/DLR) and the 
Max-Planck-Gesellschaft. We thank the ROSAT team for performing the All-Sky
Survey and producing the RASS Bright Source Catalogue. 

This work was supported in parts by the DLR under grants 50 OR 9403 5
and 50 OR 9706 8.

We gratefully acknowledge comments on and corrections to earlier 
versions of the RBS-catalogue by M. Veron, J. Retzlaff and A. Edge. 
Some identifications were kindly provided and made known 
to us before publication by N.~Bade, J.~Greiner, L.~Miller, and L.~Wisotzki.
We thank J.~Galka, M.~Gamnitzer, I.~H\"andel, G.~Hass, and D.~Rickers,
pupils from Potsdam high schools, and J.~Morfill, student at the University 
of Munich) who helped preparing the published table, 
finding charts and who performed literature searches.

This research has made use of the SIMBAD database operated at
CDS, Strasbourg, France, and the NASA/IPAC Extragalactic database (NED)
operated by the Jet Propulsion Laboratory, California Institute of 
Technology under contract with the National Aeronautics and Space 
Administration. Identification of the RASS X-ray sources was greatly
facilitated by use of the finding charts based upon the COSMOS
scans of the ESO/SERC J plates performed at the Royal Observatory 
Edinburgh and APM catalogue based on scans of the red and blue POSS plates
performed at the Institute of Astronomy, Cambridge, UK.

Based in part 
on photographic data of the National Geographic Society -- Palomar
Observatory Sky Survey (NGS-POSS) obtained using the Oschin Telescope on
Palomar Mountain.  The NGS-POSS was funded by a grant from the National
Geographic Society to the California Institute of Technology.  The
plates were processed into the present compressed digital form with
their permission.  The Digitized Sky Survey was produced at the Space
Telescope Science Institute under US Government grant NAG W-2166.


%
\addresses
\rf{A.D.~Schwope, G. Hasinger, I. Lehmann, R. Schwarz, H. Brunner, 
Astrophysical Institute Potsdam, An der Sternwarte 16, D-14482 Potsdam,
{\it aschwope,ghasinger,ilehmann,rschwarz,hbrunner@aip.de}}
\rf{J. Tr\"umper, W. Voges, 
Max-Planck-Institut f\"ur extraterrestrische Physik, 
D-85740 Garching, Germany, {\it jtrumper,whv@mpe-garching.mpg.de}}
\rf{S. Neizvestny, A. Ugryumov, Yu. Balega, 
SAO RAS, Nizhnij Arkhyz, Zelenchukskaya,
Karachaevo-Cherkesia, Russia, 357147 {\it and,balega@sao.ru}}

\end{document}